\documentclass[aps,pra,preprint,showpacs]{revtex4}
\usepackage{amsmath,graphicx}
\usepackage{epstopdf}

\begin{document}

\def\la{{\langle}}
\def\ra{{\rangle}}
\def\vep{{\varepsilon}}
\newcommand{\beq}{\begin{equation}}
\newcommand{\eeq}{\end{equation}}
\newcommand{\beqa}{\begin{eqnarray}}
\newcommand{\eeqa}{\end{eqnarray}}
\newcommand{\da}{^\dagger}

\newcommand{\wh}{\hat}
\def\la{{\langle}}
\def\ra{{\rangle}}
\def\del{\delta}
\def\om{\omega}
\def\o0{\omega_0}
\def\ov{\omega_V}
\def\os{\omega_s}
\def\O0{\Omega_0}
\def\Os{\Omega_s}
\def\Om{\Omega}
\def\do{\Delta\omega}
\def\vep{\varepsilon}
\newcommand{\iz}{\left}
\newcommand{\zi}{\right}
\newcommand{\tr}{{\rm tr}}
\newcommand{\intf}{\int_{-\infty}^\infty}
\newcommand{\into}{\int_0^\infty}

\title{Action scales for quantum decoherence and their relation 
to structures in phase space.}

\author{Daniel Alonso$^1$, S. Brouard$^2$, Jos\'e P. Palao$^2$,
R. Sala Mayato$^2$}

\affiliation{$^1$ Departamento de F\'{\i}sica Fundamental y Experimental, 
Electr\'onica y Sistemas. Universidad de La Laguna, La Laguna 38203,
Tenerife, Spain\\
$^2$ Departamento de F\'{\i}sica Fundamental II, Universidad de La Laguna,
La Laguna 38203, Tenerife, Spain \\}

\begin{abstract}

A characteristic action $\Delta S$ is defined whose magnitude determines 
some properties of the expectation value of a general quantum displacement 
operator. These properties are related to the capability of a given
environmental `monitoring' system to induce decoherence in quantum
systems coupled to it. We show that the scale for effective decoherence 
is given by $\Delta S\approx\hbar$. 
We relate this characteristic action with a 
complementary quantity, $\Delta Z$, and
analyse their connection with the main features
of the pattern of structures developed by
the environmental state in different phase space representations.
The relevance of the $\Delta S$-action scale is illustrated using both
a model quantum system solved numerically and a set of
model quantum systems for which analytical expressions for 
the time-averaged expectation value of the displacement operator 
are obtained explicitly.
\end{abstract}

\maketitle

%%%%%%%%%%%%%%%%%%%%%%%%%%%%%%%%%%%%%%%%%%%%%%%%%%%%%%%

\section{Introduction}

The superposition principle and the interference terms that generates are the
key components of the quantum formalism, and responsible for the main
differences between the quantum and classical world. The boundary between
these two worlds and the mechanisms that prevent the interference terms
from being apparent in the classical realm have been the subjects of many
theoretical and experimental studies since the very beginning of the
``quantum era''. Significant advances in the analysis and experimentation on
the interaction between mesoscopic and microscopic systems are pushing the
boundary between the two worlds. An example is the study of measurement
processes where the `monitoring' apparatus is represented by a system with an
increasingly larger number of degrees of freedom (more classical) and the
analysis of the associated disappearance of the non-diagonal terms
of the density operator of the microscopic system in some preferred 
matrix representation
\cite{zurek91,zurek2001}.
The study of the effectiveness of a given system that plays the role of an
environment or of a measurement apparatus to induce decoherence in another system
is of fundamental and practical interest. For instance, the advances in the fields
of quantum communication and quantum computation depend crucially on our ability to
manipulate entanglement \cite{popescu} and to control the capability of the 
environment
or measurement devices to induce decoherence in our qubit (pointer) system
\cite{giulini1996,wineland98}.

Many actual interactions between a two-level system {$\cal S$},
spanned by the pointer states $|+\rangle$ and $|-\rangle$, and a system
{$\cal E$} playing the role of the environment (for instance as a `monitoring'
apparatus), can be described by means of a coupling Hamiltonian of von Neumann's
form \cite{giulini1996}. 
In particular, we will use a generic term 
$\hat{V}_{\cal SE}=\left(
\left|+\right>\left<+\right|-\left|-\right>\left<-\right|\right)\,
\left({\bf c_q}\cdot{\bf \hat q}+{\bf c_p}\cdot{\bf \hat p}\right)$, 
where ${\bf \hat q}\equiv(\hat q_1,\dots,
\hat q_f)$ and ${\bf \hat p}\equiv(\hat p_1,\dots,\hat p_f)$ are position and
momentum operators for an environmental system with $f$ degrees of
freedom ($\left[\hat q_j,\hat p_j\right]=i\hbar$, $j=1,\dots,f$).
The coefficients ${\bf c_q}\equiv(c_q^{(1)},\dots,c_q^{(f)})$ and 
${\bf c_p}\equiv(c_p^{(1)},\dots,c_p^{(f)})$ 
characterise the strength of the coupling.
The reduced density operator describing the state of the system ${\cal S}$
after its coupling with the environment during a time interval $\delta t$ is
given by  
\beq\label{eq:reduced}
\hat{\rho}_{\cal S}\,=\,|\alpha|^2 \left|+\right>\left<+\right|+|\beta|^2
\left|-\right>\left<-\right|\\
+\left(\alpha\beta^*\left<\psi_-|\psi_+\right>\left|+
\right>\left<-\right|+{\rm H.c.}\right)\,,
\eeq
where $\left|\psi_\pm\right>\equiv\hat D(\mp {\bf c_p}\delta t,
\mp {\bf c_q}\delta t)\left|\psi\right>$, $\hat{D}({\bf \delta q},
{\bf \delta p})\equiv\exp\{i({\bf \hat{p}}\cdot{\bf \delta q}+
{\bf \hat{q}}\cdot{\bf \delta p})/\hbar\}$,
${\bf \delta q}$ and ${\bf \delta p}$ are displacement vectors in
$f$-dimensional spaces, and H.c. denotes the Hermitian conjugate
of the preceding term in the equation.
The states of the environmental and two-level systems immediately 
prior to the interaction are 
$\left|\psi\right>$ and $\left|\chi\right>\equiv\alpha
\left|+\right>+\beta\left|-\right>$ respectively.
We have assumed that the coupling strength
is large enough so that the evolution 
induced by each system Hamiltonian ($\hat{H}_{\cal S}$ and $\hat{H}_{\cal E}$)
can be neglected during the interaction time $\delta t$ \cite{giulini1996}.
Despite the simplicity of the model considered, it contains
the basic elements relevant to our discussion. 

Eq. (\ref{eq:reduced}) relates the value of the non-diagonal term of the reduced
density matrix of ${\cal S}$ in the preferred basis $\{|+\rangle,|-\rangle \}$
to the mean value of a displacement operator over the state 
$|\psi\rangle$ of system $\cal E$
since $\left<\psi_-|\psi_+\right>=\left<\psi\right|\hat D(-2{\bf c_p}\delta t,
-2{\bf c_q}\delta t)\left|\psi\right>$. Therefore the capability of $\cal E$
to induce decoherence in $\cal S$ through the coupling term $\hat{V}_{\cal SE}$ is
characterised by 
\begin{equation}\label{eq:overlapstate}
C_{\psi}({\bf \delta q},{\bf \delta p})
\equiv
\langle\psi|\hat{D}({\bf \delta q},{\bf \delta p})|\psi \rangle
\,=\,
e^{i {\bf \delta q\cdot\delta p} /2 \hbar} \int d^f\! q
\, e^{i{\bf q\cdot\delta p} /\hbar}\,\psi^*({\bf q})\,\psi({\bf q+\delta q})\,,
\end{equation}
where $\psi({\bf q})\equiv\langle {\bf q}|\psi\rangle$,
${\bf \delta q}=-2{\bf c_p}\delta t$, ${\bf \delta p}=-2{\bf c_q}\delta t$,
and $d^f\!q$ ($d^f\!p$) is the $f$-dimensional differential element of volume
in positions (momenta). 
All integrals in this paper run over the entire available volume.
Complete decoherence is reached whenever the
two states $|\psi_+\rangle$ and $|\psi_-\rangle$ are orthogonal to each other;
in other words, when $C_\psi=0$.
At this point, it is important to characterise
the scale for which displacements $({\bf \delta q},{\bf \delta p})$ in phase
space will produce a significant decay of this expectation value of $\hat D$.
The main subject of our interest
is to find an action scale associated to the effectiveness 
of system $\cal E$ to induce decoherence in system $\cal S$, 
and to describe its dependence with the particular environmental 
state.

This question has been previously studied by Zurek \cite{zurek2001} 
by means of the Wigner phase space distribution associated to the state 
$|\psi\ra$ \cite{Wigner1},
\begin{equation}\label{eq:wigner}
W_\psi({\bf q},{\bf p})
\,=\,
\frac{1}{(2 \pi \hbar)^f}  
\int d^f\! {q}'\, e^{i{\bf q}'\cdot{\bf p}/\hbar}\,\psi({\bf q}-{\bf q}'/2)\,
\psi^*({\bf q}+{\bf q}'/2)\,.
\end{equation}
In particular, Moyal's formula \cite{Moyal} 
\begin{equation}\label{eq:wigwig}
\left|C_{\psi}({\bf \delta q},{\bf \delta p})\right|^2
\,=\,
(2\pi\hbar)^f\,\int\,d^f\! q\,d^f\!p\,
W_{\psi}({\bf q},{\bf p})\,W_{\psi}({\bf q+\delta q},{\bf p+\delta p})\,
\end{equation}
was used to analyse the behaviour of the overlap $|C_\psi|^2$ with ${\bf \delta q}$
and ${\bf \delta p}$. 
The choice of the Wigner phase space 
distribution was motivated by this simple expression for 
the scalar product between $|\psi_+\rangle$ and $|\psi_-\rangle$.
In Ref. \cite{zurek2001} Zurek showed that for a given
time-dependent quantum chaotic system in one dimension ($f=1$) confined to a
phase space volume characterised by the classical action $A$, the Wigner
distribution associated to the state develops in time
a spotty random structure on the scale $\hbar^2/A$. 
Using Eq. (\ref{eq:wigwig}) he argued that $|C_{\psi}|^2\approx 0$ 
for phase space displacements on the scale of
the smallest structure of the Wigner distribution $W_\psi(q,p)$. The basis for
this result are: (a) Displacements characterised by $\delta q\delta p\approx
\hbar^2/A$ produce a significant decrease on the value of the integral in Eq.
(\ref{eq:wigwig}) due to the destructive interference between $W_\psi(q,p)$
and $W_\psi(q+\delta q,p+\delta p)$, and (b) the random distribution of the
patches in the structure appearing in the Wigner function associated to 
such system states prevents the presence of recurrences in the value 
of the overlap.

Jordan and Srednicki \cite{jordan2001} extended the analysis in Ref.
\cite{zurek2001} to systems with an arbitrary number of degrees of freedom
by using
\begin{equation}\label{eq:overlapwig}
C_{\psi}({\bf \delta q},{\bf \delta p})
\,=\,
\int\,d^f\! q\,d^f\! p\, 
e^{i({\bf p\cdot\delta q} +{\bf q\cdot\delta p})/\hbar}\,W_\psi({\bf q},{\bf p})\,.
\end{equation}
This equation establishes a relation between the small-scale
(large-scale) structure of $W_\psi$ in the variables $({\bf q},{\bf p})$ and
the large-scale (small-scale) structure of $C_{\psi}$ 
in the variables $({\bf \delta p},{\bf \delta q})$. 
Analysing a two-dimensional billiard and a gas of $N$ hard spheres
in a three-dimensional box (assuming the Berry-Voros conjecture
\cite{berry_voros} in both cases) 
they concluded that for systems with a small number of degrees 
of freedom, displacements
$\delta q_i\approx L_i$ and $\delta p_i\approx P_i$ are needed to
avoid oscillations in the overlap, where $L_i$ and $P_i$ are typical
classical values of the position $q_i$ and momentum $p_i$ respectively
($i=1,\dots,f$).
This means that displacements of the order of the size of the state support
are needed to guarantee orthogonality in the general case. However, for
systems with a large number of degrees of freedom they found that the
conclusions in Ref. \cite{zurek2001} remain valid, supporting the idea
that a larger number of degrees of freedom increases the
effectiveness in causing decoherence.
Some care must be taken when relating the results in Refs.\cite{zurek2001} and
\cite{jordan2001} since in principle the Berry-Voros conjecture is not valid
for the system analysed by Zurek in Ref. \cite{zurek2001} and 
the dependence of the overlap with the displacement could have 
qualitatively different features.

In this work we characterise the behaviour of $C_{\psi}$ using
a quantity $\Delta S$, with units of action, associated to the
displacement $(\delta{\bf q},\delta{\bf p}$).
A formal series expansion of $\hat D$ will allow us to identify
the scale in the action $\Delta S(\delta{\bf q},\delta{\bf p})$
for which the overlap decreases significantly for any quantum 
system, irrespective of the number of degrees of freedom.
This scale is manifested in the size of the structures present in
the distribution associated to the state in 
some phase space representations, but they do not necessarily coincide.

The paper is organised as follows. In Sec. II we define the 
characteristic action $\Delta S$ and determine the 
scale relevant for the decay of the overlap. 
In Sec. III we establish the relation between $C_\psi$
and the structure of the distribution associated to the
state in an arbitrary phase space representation.
The next two sections are devoted to studying in detail
the dependence of $C_\psi$ on $\Delta S$ 
for states of particular quantum systems.
Sec. IV considers a system with a time-dependent
Hamiltonian whose classical counterpart exhibits chaos.
In Sec. V we analyse the case of
non-linear systems with a confining potential and discrete 
spectrum. In this case the main features of $C_\psi$ can be
obtained from time average properties of the state evolution. 
We will focus on quantum systems with time-independent
Hamiltonian for which analytical models are worked out by 
using the Berry-Voros conjecture \cite{berry_voros}.
Finally in Sec. VI the main results of this work are
discussed.

%%%%%%%%%%%%%%%%%%%%%%%%%%%%%%%%%%%%%%%%%%%%%%%%%%%%%%%%%%%%%%%%%%%%%%%%%%%%%%
\section{Characteristic action scales for the decay of 
$C_{\psi}({\bf\delta q},{\bf \delta p})$}

A displacement operator $\hat D({\bf \delta q},{\bf \delta p})$ acting on the
state of an $f$-dimensional quantum system ${\cal E}$, that describes an
environment or a `monitoring' apparatus, can be written as 
\begin{equation}\label{eq:DandS}
\hat D({\bf \delta q},{\bf \delta p})
\,=\,
e^{i ({\bf \hat p\cdot\delta q}+{\bf \hat q\cdot\delta p})/\hbar}
\equiv e^{i\hat S({\bf \delta q},{\bf \delta p})/\hbar}\,.
\end{equation}
The main features of $|C_\psi|^2$ are therefore related to the 
fluctuation properties of the operator $\hat{S}({\bf \delta q},{\bf \delta p})$, 
since the expectation value of $\hat{D}$ equals the characteristic 
function of $\hat S$ (see Eq. (\ref{eq:overlapstate})).

A formal expansion of $\hat D$ in terms of $\hat S$ gives 
\begin{equation}\label{eq:overlapS}
C_{\psi}({\bf \delta q},{\bf \delta p})
\,=\, 
1 + 
\frac{i}{\hbar} \langle \hat S \rangle_{\psi}-\frac{1}{2 \hbar^2}
\langle \hat S^2 \rangle_{\psi}
+ {O}\left(\frac{s^3\delta^3}{\hbar^3}\right)\,,
\end{equation}
and for the overlap,
\begin{equation}\label{eq:sqmodulusS}
|C_{\psi}({\bf\delta q},{\bf\delta p})|^2
\,=\,1 - \frac{1}{\hbar^2}
\Big(\langle \hat S^2 \rangle_{\psi}-\langle \hat S \rangle_{\psi}^2
\Big)
+O\left(\frac{s^4\delta^4}{\hbar^4}\right)\,.
\end{equation}
We denote by $\delta^n$ general products of $n$ components of the vectors
${\bf\delta q}$ and ${\bf \delta p}$,
and by
$s^n$ terms of the form $\prod_{k=1}^{m}\langle\hat{O}_k\rangle_\psi$, 
where $\hat{O}_k$ is the product of $g_k$ operators ${\hat q}$ and
${\hat p}$, with the condition $\sum_{k=1}^{m} g_k=n$.
The characteristic action
\begin{equation}
\Delta S (\delta{\bf q},\delta{\bf p})\equiv \,
\sqrt{\langle\hat S^2\rangle_{\psi}-\langle\hat S\rangle_{\psi}^2}\,
\end{equation}
controls the decay of the overlap for 
sufficiently small values of ${\bf\delta q}$ and ${\bf\delta p}$.
Eq. (8) suggests that displacements $(\delta{\bf q},\delta{\bf p})$
for which $\Delta S$ is small compared to $\hbar$
do not lead generally to an important decay of  
$|C_{\psi}|^2$. 
In other words, displacements leading to $\Delta S$ of 
the order or larger than $\hbar$ are needed for the states
$|\psi_+\rangle$ and $|\psi_-\rangle$ to be orthogonal.
Therefore $\Delta S\approx\hbar$ establishes the scale for 
the action involved in displacements of the environmental
state that could induce significant decoherence in system 
$\cal S$.
For the case of Gaussian fluctuations of the operator $\hat{S}$, 
the only relevant fluctuation is $\Delta S$. In a more general 
situation higher order fluctuations may play a role in the
particular features of the decay of $|C_\psi|^2$, nonetheless
the $\Delta S$-action scale is generally expected to be a good 
measure for the decoherence process.  
The rest of the paper will provide additional arguments
for this interpretation of the scale associated to the
quantity $\Delta S$.

To be more specific, let us write 
$(\Delta S)^2$ in terms of $({\bf\delta q},{\bf\delta p})$,
\begin{eqnarray}\label{eq:completa}
(\Delta S)^2&=&\sum_{i=1}^f \sum_{j=1}^f \Big[
(\la \hat q_i \hat q_j\ra_\psi 
- \la\hat q_i\ra_\psi \la\hat q_j\ra_\psi)\delta p_i\delta p_j +
(\la \hat p_i \hat p_j\ra_\psi 
- \la\hat p_i\ra_\psi \la\hat p_j\ra_\psi)\delta q_i\delta q_j
\nonumber\\
&+&(\la \hat q_i \hat p_j\ra_\psi 
- \la\hat q_i\ra_\psi \la\hat p_j\ra_\psi)\delta p_i\delta q_j +
(\la \hat p_i \hat q_j\ra_\psi 
- \la\hat p_i\ra_\psi \la\hat q_j\ra_\psi)\delta q_i\delta p_j
\Big]\,,
\end{eqnarray}
or
\begin{equation}\label{eq:multidim}
(\Delta S)^2\,=\,
{\bf\delta p}^{\bf T}\boldsymbol{\gamma}^{qq}{\bf\delta p} +
{\bf\delta q}^{\bf T}\boldsymbol{\gamma}^{pp}{\bf\delta q} +
{\bf\delta p}^{\bf T}\boldsymbol{\gamma}^{qp}{\bf\delta q} +
{\bf\delta q}^{\bf T}\boldsymbol{\gamma}^{pq}{\bf\delta p}\,,
\end{equation}
where we have introduced the matrices
$\boldsymbol{\gamma}^{AB}_{ij}\equiv\la\hat A_i\hat B_j\ra_\psi-
\la\hat A_i\ra_\psi\la\hat B_j\ra_\psi$, and ${\bf a}^{\bf T}$ 
denotes the transposed of the vector ${\bf a}$.
To gain some insight into the meaning of this quantity, 
we will consider $(\Delta S)^2$ for the one-dimensional case, 
\beq
(\Delta S)^2\,=\,({\sigma}_q{\delta p})^2
+({\sigma}_p{\delta q})^2
+(\langle{\hat q\hat p}+{\hat p\hat q}\rangle_{\psi}
\,-\,2\langle {\hat q}\rangle_{\psi}
\langle{\hat p}\rangle_{\psi})
{\delta q\delta p}\,,
\eeq
where ${{\sigma}_q}$ and ${{\sigma}_p}$ are the root-mean-square
deviations of ${\hat q}$ and ${\hat p}$ respectively. 
To continue with our discussion, a rotation in phase space is made,
so that the term 
$(\langle{\hat q\hat p}+{\hat p\hat q}\rangle_{\psi}-2\langle {\hat q}
\rangle_{\psi}\langle{\hat p}\rangle_{\psi})$ 
in the previous equation is zero, and $\Delta S$ is given, 
in terms of the new phase space variables, by
\beq\label{eq:rotation}
(\Delta S)^2\,=\,({\sigma}_{\tilde q}{\delta {\tilde p}})^2\,+\,
({\sigma}_{\tilde p}{\delta {\tilde q}})^2\,,
\eeq
where ${\sigma}_{{\tilde q}}$ (${\sigma}_{{\tilde p}}$) gives the support of the
state in the variable ${\tilde q}$ (${\tilde p}$). A classical action 
$A\equiv{\sigma}_{\tilde q}{\sigma}_{\tilde p}$ can be associated to the state of
the system. 
Eq. (\ref{eq:rotation}) implies that
displacements such that 
${\sigma}_{\tilde q}\,{\delta\tilde p}\approx\hbar$ or 
${\sigma}_{\tilde p}\,{\delta\tilde q}\approx\hbar$ give
$\Delta S\gtrsim\hbar$,
and the main point of our analysis is that they
also lead in general to a significant variation of $|C_{\psi}|^2$,
irrespective of the value of action $A$.
In this sense values of the order or larger than
$\hbar$ of the $\Delta S$-action scale are always needed
for this environmental system to induce decoherence.
It is possible to define other relevant
quantities with units of action. For instance,
values of $\Delta Z\equiv\delta\tilde{q}\delta\tilde{p}$
leading to a significant decrease of
the overlap are related to the size of the structure of the 
distribution associated to the state 
in some particular phase space representations \cite{zurek2001}.
For the displacements discussed above
$\Delta Z\approx\hbar^2/A$, and if $A>>\hbar$
the result that sub-Planck displacements on the 
$\Delta Z$-action scale are relevant for the decoherence
process induced by $\cal E$ comes naturally.
Coming back to the multi-dimensional case, 
when the dimension of the problem increases more terms will 
contribute to $(\Delta S)^2$ in Eq. (\ref{eq:completa}),
and smaller displacements in each variable are needed to reach the 
threshold $\Delta S\approx\hbar$, leading to the result that a larger 
number of degrees of freedom will favour the decoherence 
process \cite{jordan2001}.

To illustrate the difference between $\Delta Z$- and $\Delta S$-action 
scales we consider a general Gaussian state in one dimension
\beq
\psi(q)=\left(\frac{2 z_R}{\pi|z|^2}\right)^{1/4} 
e^{ip_0q/\hbar} e^{-(q-q_0)^2/z},
\eeq
with $z\equiv z_R+iz_I$, $z_R=(\hbar/\sigma_p)^2$, and $z_I=z_R\sqrt{4
\sigma_q^2\sigma_p^2-\hbar^2}/\hbar$. Straightforward calculations lead
to the exact expression
\beq\label{eq:cforgaussian}
|C_\psi(\delta q,\delta p)|^2
\,=\,
\exp\left[-(\Delta S)^2/\hbar^2\right]
\,,
\eeq
where
\beq\label{eq:dsforgaussian}
(\Delta S)^2=(\sigma_p\delta q)^2+
(\sigma_q\delta p)^2+\hbar\sqrt{\left(\frac{2\delta p\delta q\sigma_p
\sigma_q}{\hbar}\right)^2-\left(\delta p\delta q\right)^2},
\eeq
in terms of the first two moments of $\hat S$, as expected 
for a Gaussian wavefunction.
Eq. (\ref{eq:cforgaussian}) shows that values of 
$\Delta S\gtrsim\hbar$ are 
needed to obtain a significant decrease of the overlap $|C_\psi|^2$. 
If we now choose, for instance, particular values of the widths 
$\sigma_q$ and $\sigma_p$ so that the Gaussian state is much 
narrower in coordinate than in momentum space, say 
$\sigma_q\simeq\sqrt{\hbar}/10$ and  $\sigma_p\simeq 10\sqrt{\hbar}$ 
(in arbitrary units), it is clear that a displacement 
$(\delta q,\delta p)=(\sqrt{\hbar}/2,\sqrt{\hbar}/2)$ 
will take the shifted Gaussian completely away from the initial one.
The different actions associated to that same displacement 
are $\Delta S \approx 5 \hbar$ and $\Delta Z = \hbar/4$,
corresponding to over-Planck and sub-Planck values
respectively.
%

%%%%%%%%%%%%%%%%%%%%%%%%%%%%%%%%%%%%%%%%%%%%%%%%%%%%%%%%%%%%%%%%%%%%%%
%%%%%%%%%%%%%%%%%%%%%%%%%%%%%%%%%%%%%%%%%%%%%%%%%%%%%%%%%%%%%%%%%%%%%%%

\section{Sub-Planck structures in phase space distributions.}
\label{sec:structures} 

The behaviour of the overlap $|C_\psi|^2$
with $(\delta{\bf q},\delta{\bf p})$
can be alternatively studied through
the distribution associated to the state in
different phase space representations. 
In this section we will derive the relation between the 
overlap and the action $\Delta S$ using a wide class
of quantum quasi-probability distributions
$F({\bf q},{\bf p};\chi)$ \cite{napoles}, 
the Wigner \cite{Wigner1} and Husimi
\cite{Cartwright76} functions being nothing but particular cases.
The choice among the $F$ functions associated to the same quantum state
of a system, or, equivalently, the selection of a particular representation
(given by function $\chi$), is similar to the choice of a convenient set of
coordinates \cite{Cohen1,Cohen2,SPM}.
Within this framework, the expectation value of any operator
$\wh G({\bf\hat q},{\bf\hat p})$ is written as the phase space integral
\beq\label{eq:meangen}
\la\wh G({\bf\hat q},{\bf\hat p})\ra_\psi\,=\,
\int d^f\!q\,d^f\!p\,\,
F_\psi({\bf q},{\bf p};\chi)\,g({\bf q},{\bf p};\chi)\,,
\eeq
where $F_\psi({\bf q},{\bf p};\chi)$ is obtained from the quantum state
$|\psi\ra$ as
\beq\label{eq:efegen}
F_\psi({\bf q},{\bf p};\chi)=\frac{1}{(2\pi)^{2f}}
\int d^f\!{\theta}\, d^f\!{\tau}\, d^f\!{u}\,\,
\chi(\boldsymbol{\theta},\boldsymbol{\tau})
\bigg <{\bf u}+\frac{\boldsymbol{\tau}\hbar}{2}\bigg|\psi\bigg >\bigg <\psi
\bigg|{\bf u}-\frac{\boldsymbol{\tau}\hbar}{2}\bigg >
e^{-i[\boldsymbol{\theta}\cdot({\bf q}-{\bf u})+\boldsymbol{\tau}\cdot{\bf p}]}\,.
\eeq
The Wigner and Husimi functions, for instance, are obtained by replacing
$\chi(\boldsymbol{\theta},\boldsymbol{\tau})=1$ and $\chi(\boldsymbol{\theta},
\boldsymbol{\tau})=\exp\{-\frac{\hbar}{4}[(\boldsymbol{\tau}\lambda)^2+
(\boldsymbol{\theta}/\lambda)^2]\}$ respectively.
The function $g({\bf q},{\bf p};\chi)$ is the {\em image} 
of the operator $\wh G$ in phase space according to the kernel function $\chi$
\cite{SPM},
\beq\label{eq:imageop}
g({\bf q},{\bf p};\chi)\,=\,\left(\frac{\hbar}{2\pi}\right)^f
\int d^f\!\theta\, d^f\!\tau\, d^f\!u\,\,
\frac{1}{\chi(\boldsymbol{\theta},\boldsymbol{\tau})}
\bigg<{\bf u}-\frac{\boldsymbol{\tau}\hbar}{2}\bigg|\hat{G}
\bigg|{\bf u}+\frac{\boldsymbol{\tau}\hbar}{2}\bigg>
e^{i[\boldsymbol{\theta}\cdot({\bf q}-{\bf u})+
\boldsymbol{\tau}\cdot{\bf p}]}\,,
\eeq
and it is not necessarily equal to the classical magnitude. 
In particular, the expectation value of the displacement operator 
$\hat D$ can be written as the phase space average
\begin{eqnarray}\label{Uni}
C_{\psi}({\bf\del q},{\bf\del p})\,&=&\,
\int d^f\!q\,d^f\!p\,\, F_\psi({\bf q},{\bf p};\chi)\,
d({\bf q},{\bf p};\chi)\,\nonumber\\
&=&\,
\frac{1}{\chi({\bf\del p}/\hbar,{\bf\del q}/\hbar)}
\int d^f\! q\,d^f\! p\,\,
F_\psi({\bf q},{\bf p};\chi)\,
e^{i({\bf q\cdot\del p}+{\bf p\cdot\del q})/\hbar}\,,
\end{eqnarray}
where the function $d({\bf q},{\bf p};\chi)$ was obtained 
by integrating the r.h.s. of Eq. (\ref{eq:imageop}) with 
$\hat{G}$ replaced by $\hat{D}$.
Notice that Eq. (\ref{eq:overlapwig}) is a particular case of 
Eq. (\ref{Uni}) for which the Wigner function has been chosen 
as the distribution associated to the state, 
$C_{\psi}({\bf \del q},{\bf \del p})$ being 
equal to the Fourier transform of $W_{\psi}$.
Eq. (\ref{Uni}) can be used to understand the relation between $C_{\psi}$ and
$\Delta S$ from the point of view of the phase space distribution $F_\psi$.
On one hand, if the exponential factor does not vary significantly
over the support of $F_\psi({\bf q},{\bf p};\chi)$,
which occurs for small enough values of ${\bf\delta q}$ 
and ${\bf\delta p}$, 
then the overlap will
only differ slightly from the normalisation integral of the
original distribution, 
$\int d^f\! q\,d^f\! p\,\,F_\psi({\bf q},{\bf p};\chi)=1$,
leading in general to a small decrease of the 
function $|C_{\psi}|^2$. (Notice that $\chi(0,0)=1$ is needed to 
guarantee that $F_\psi({\bf q},{\bf p};\chi)$ is normalised
to one \cite{Cohen2}.) The condition for these variations not 
to be significant is equivalent to
the condition that the root-mean-square deviation of $\hat S$, $\Delta S$, 
is smaller than $\hbar$. On the other hand, to obtain significant 
decay of the overlap, rapid oscillations, with
$\Delta S$ at least of the order of $\hbar$, are needed.
Due to the properties of the Fourier transform, and since the value of 
$\chi({\bf\del p}/\hbar,{\bf\del q}/\hbar)$ 
is close to one for small enough ${\bf \del q}$
and ${\bf \del p}$, the initial decay of $|C_\psi|^2$ with 
the displacement is related to the large scale structure of the 
distribution $F_\psi$, that depends mainly on the size of the state 
support in phase space. However, the detailed behaviour of $|C_\psi|^2$
with arbitrary displacements, and, in particular, qualitative features 
like oscillations, will depend on the state under study.

A different question is how sub-Plank structures emerge in some
phase space distributions associated to the state and how they 
are related to the main features of $C_\psi$. 
For kernel functions such that 
$|\chi(\boldsymbol{\theta},\boldsymbol{\tau})|=1$, the corresponding 
distributions verify \cite{tw1}
\begin{equation}\label{eq:scalar_product_k1}
\left|C_{\psi}({\bf \delta q},{\bf \delta p})\right|^2
\,=\,
(2\pi\hbar)^f\,\int\,d^f\! q\,d^f\!p\,
{F}_{\psi}({\bf q},{\bf p})\,
{F}_{\psi}({\bf q+\delta q},{\bf p+\delta p})\,.
\end{equation}
(Notice that Eq. (\ref{eq:wigwig}) is a particular case of Eq. 
(\ref{eq:scalar_product_k1}).)
For these representations, the fact that a given displacement 
leads to $|C_\psi|^2\approx 0$
is manifested in a complex structure of the distribution
$F_\psi$ on the scale 
of the $\Delta Z$-action for that displacement, as 
pointed out in Ref. \cite{zurek2001} for the Wigner distribution.
When displacements with $\Delta Z<<\hbar$ 
lead to small values of $|C_\psi|^2$, the distribution $F_\psi$ 
will show a complex structure at sub-Planck scales. 
This result is a consequence of the particular choice of the
phase space distribution. For the same state, the Husimi distribution 
(obtained by smoothing the Wigner function, so eliminating the 
sub-Planck scale structure) will lead to the same overlap $|C_\psi|^2$.
This is not surprising since the overlap depends on the state, 
and that dependence is manifested in different ways for different 
phase space representations.

%%%%%%%%%%%%%%%%%%%%%%%%%%%%%%%%%%%%%%%%%%%%%%%%%%%%%%%%%%%%%%%%%%%%%%

\section{A one-dimensional time-dependent environmental system}
\label{sec:1dtd}

In this section we will analyse the dependence of the overlap
$|C_{\psi}|^2$ on the action $\Delta S(\delta q,\delta p)$ in
the context of a particular one-dimensional model for the
environmental system $\cal E$, described by Hamiltonian
\begin{equation}\label{eq:HamZurek}
\hat H_{\cal E}\,=\,\frac{\hat p^2}{2m}
-\kappa\cos\left(\hat q-l\sin t\right)+\frac{1}{2}a\hat q^2\,.
\end{equation}
This quantum model system has been previously used in the context of 
decoherence \cite{zurek2001,KJZ02}, and describes a particle of mass $m=1$
(arbitrary units are used throughout) confined by a harmonic
potential that is perturbed by a spatially and temporally periodic
term. For the parameter values used in this work, $\kappa=0.36$,
$a=0.01$, $l=3.8$, and $\hbar=0.16$, the motion in the 
classical counterpart of this system exhibits
a chaotic character \cite{KZZ02}.

To prepare the state of the environmental system prior to the
interaction, we let a given initial state evolve until
preparation time $T$, when it is coupled to the pointer
system $\cal S$.
The coupling strength is assumed to be high enough so that the
two-systems evolution can be followed as described in the
introduction, i.e., by neglecting any contribution coming from
the dynamics induced by Hamiltonian (\ref{eq:HamZurek}) 
during the interaction time.
This approach allows us to discuss the values of actions involved
in the decoherence process induced on system $\cal S$ in terms of
general displacements in phase space, irrespective of the detailed
values of the coupling constants and interaction times \cite{commentprl}.

As the initial state ($T=t=0$) for the preparation process we have
chosen a coherent state 
$|\alpha\ra\,=\,e^{-|\alpha
|^2/2}\,\sum_{n=0}^\infty\frac{\alpha^n}{\sqrt{n!}}\,|n\ra\,$ 
of the harmonic oscillator 
$\hat H_{OA}=\hat p^2/(2m)+a \hat q^2/2$,  
where $|n\ra$ is the eigenstate of $\hat H_{OA}$ with energy 
$(n+1/2)\hbar\sqrt{a/m}$.
(We have checked other possible initial states 
obtaining qualitatively similar results.) 
Its time propagation under
Hamiltonian (\ref{eq:HamZurek}) has been obtained by means of 
the split-operator method \cite{FFS82}.
%

%%%%%%%%%%%%%%%%%%%%%%%%%%%%%%%%%
\begin{figure}
\includegraphics[width=10cm]{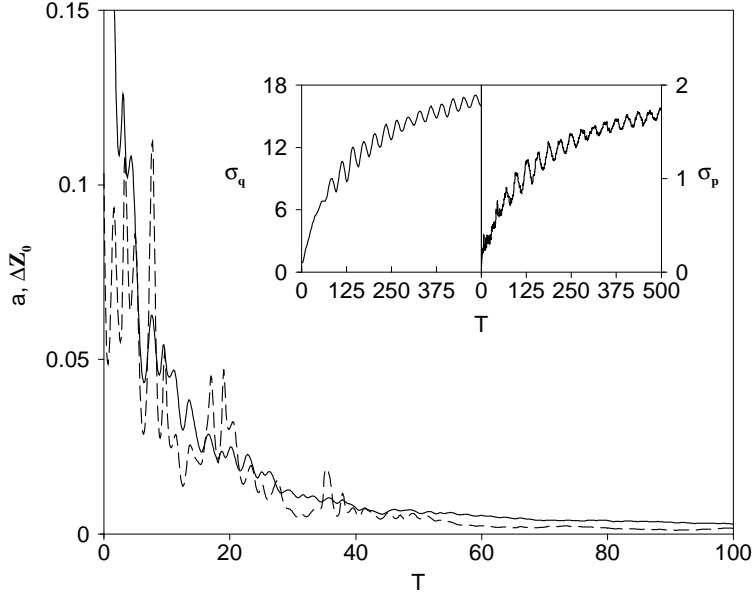}
%\vskip0.5cm
\caption[]{Action $a\equiv\hbar^2/(\sigma_q\sigma_p)$ as a function of
the preparation time $T$ (solid line) for the initial coherent
state with $\alpha=5\,i$.
The action $\Delta Z_0$ needed for a displacement in the direction 
$\delta q\simeq 6.8\delta p$ to reduce the value of $|C_\psi|^2$
to $0.5$ is also shown (dashed line). The inset shows the dependence
of the widths $\sigma_q$ (left) and $\sigma_p$ (right) with 
the preparation time. (Notice the different scales in the 
vertical axis for each case.) Arbitrary units are used.}
\label{fig:fig1}
\end{figure}
%%%%%%%%%%%%%%%%%%%%%%%%%%%%%%%%%%

As time $T$ increases, the state spreads in coordinate as well as
in momentum space through the available phase space as shown in
the insets of Fig. \ref{fig:fig1}. 
To characterise this dynamics the quantity
\begin{equation}\label{eq:definitiona}
a\equiv\frac{\hbar^2}{\sigma_q\sigma_p}\,
\end{equation}
is used (see Fig. \ref{fig:fig1}). It shows a rapid initial decay (until time
$T\approx 20$), followed by a much slower decrease for longer times.
The behaviour of $a$ for small preparation time $T$
is related to the fast initial increase of the widths 
$\sigma_q$ and $\sigma_p$.
The variation of $a$ for longer times is mainly due to the 
time-dependent term in the Hamiltonian. Should not be for the presence
of this time-dependent term $a$ would not decrease beyond a certain
minimum value related to the maximum position and momentum
widths compatible with a fixed mean system energy.

%%%%%%%%%%%%%%%%%%%%%%%%%%%%%%%%%
\begin{figure}
\includegraphics[width=10cm]{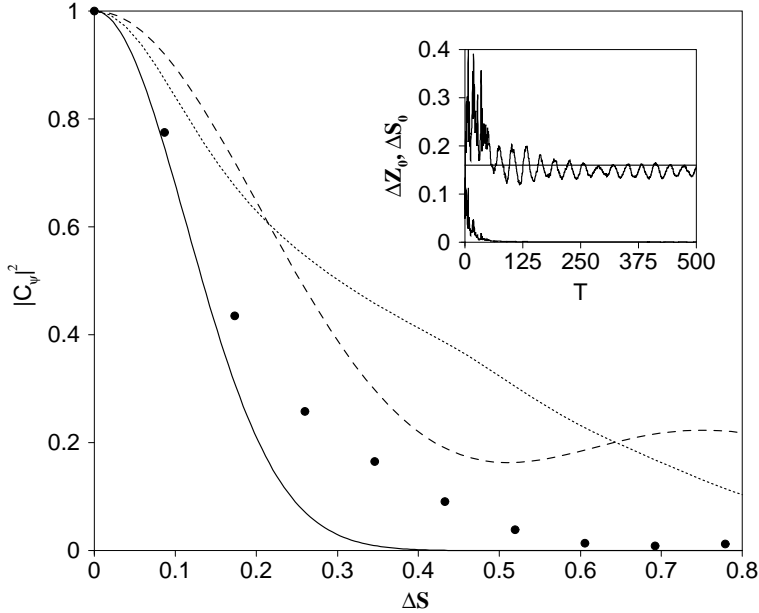}
%\vskip0.5cm
\caption[]{$|C_\psi|^2$ as a function of the action 
$\Delta S(\delta q,\delta p)$ 
in the direction $\delta q\simeq 6.8\delta p$ in
phase space and for different preparation times:
$T=0$ (solid line), $T=10$ (dashed line), $T=20$ (dotted line), and
$T=500$ (circles).
(Same initial state as in Fig. \ref{fig:fig1}.)
The inset shows the actions $\Delta S_0$ (upper curve) and $\Delta Z_0$
(lower curve) needed for the displacement to 
reduce $|C_\psi|^2$ to the value $0.5$ versus the preparation time $T$.
The straight line shows the action $\hbar$ for reference. Arbitrary units
are used.}
\label{fig:fig2}
\end{figure}
%%%%%%%%%%%%%%%%%%%%%%%%%%%%%%%%%%

Fig. \ref{fig:fig2} shows $|C_\psi|^2$ versus $\Delta S$ for different 
preparation times and for a given direction in phase space.
(The results for any other direction show the same 
qualitative features.)
The different curves, corresponding to different preparation
times, decay in the same $\Delta S$-scale.
To emphasise this result we represent in the inset the value
of $\Delta S$ needed to obtain  $|C_{\psi}|^2=0.5$ versus
$T$. The $\Delta S$-action values for any preparation time 
are of the order of $\hbar$, supporting $\Delta S\approx\hbar$
as a relevant scale for the studied decoherence process.
(Notice that the apparent convergence of $\Delta S_0$ to 
a value close to $\hbar$ is only a consequence of the chosen 
value for $|C_{\psi}|^2$.)

The values of $\Delta Z$ for which $|C_{\psi}|^2=0.5$ are also
shown in Fig. \ref{fig:fig1} and Fig. \ref{fig:fig2}.
After some time $T\approx 5$,
the action $a$ sets the scale of the random structure developed 
in the distribution associated to the states of this system in some
phase space representations, for example in the Wigner function 
\cite{zurek2001}. Taking into account the discussion 
below Eq. (\ref{eq:scalar_product_k1}), 
the action $\Delta Z$ for displacements producing 
a significant decrease of $|C_\psi|^2$ will be of the 
order of the action $a$ after $T\approx 5$, 
as shown in Fig. \ref{fig:fig1}.

%%%%%%%%%%%%%%%%%%%%%%%%%%%%%%%%%%%%%%%%%%%%%%%%%%%%%%%%%%%%%%%%%

\section{Non-linear confined environmental systems}
\label{sec:nonlinear}

For now on, a different model for the environmental system 
$\cal E$ will be considered, that of a time-independent Hamiltonian,
$\hat{H}_{NL}$,
with a non-linear confining potential and a discrete energy spectrum.
Under certain assumptions, this model will allow us 
to obtain analytical expressions for the overlap between
the states $|\psi_+\rangle$ and $|\psi_-\rangle$. Instead
of considering a particular environmental state $|\psi\rangle$,
obtained after some fixed preparation time $T$, we will study
the dependence of the overlap averaged over
the preparation time on an averaged $\Delta S$-action.
For non-linear confined systems, the main features 
of this stationary description
can be associated to all the states prepared from a 
given initial one, provided that their preparation time is 
long enough.
In the first part of this section we will determine the 
stationary properties of $C_{\psi}$ relevant to our discussion.
We will assume that the states are prepared from a given 
$|\psi(T=0)\rangle$, and make use of the Wigner distribution 
in phase space associated to them.
Although the procedure and the results are independent of the
choice of a particular phase space representation,
the use of the Wigner distribution will allows us to 
extend our analysis afterwards for systems for which 
the Berry-Voros conjecture is valid.

\subsection{Stationary properties of the overlap}

In the basis of eigenstates of the Hamiltonian $\hat{H}_{NL}$, which 
will be assumed to have, for simplicity, a non-degenerate spectrum,
the wave function at preparation time $T$ is given by
\begin{equation}\label{psi}
\psi({\bf q},T)=\sum_n c_n e^{-i E_n T/\hbar} \varphi_n({\bf q}),
\end{equation}
where $\hat H\varphi_n({\bf q})=E_n \varphi_n({\bf q})$ and 
$c_n=\int dq \, \varphi_n^*({\bf q}) \psi({\bf q},0)$.
The Wigner distribution is obtained introducing expansion 
(\ref{psi}) into Eq. (\ref{eq:wigner}). Splitting the result into
time-independent and time-dependent terms,
\begin{eqnarray}
W_\psi({\bf q},{\bf p},T)&=& 
\sum_n |c_n|^2 W_{\varphi_n}({\bf q},{\bf p})\\
&+& \sum_{n\ne m}c_n c^*_m e^{-i (E_n-E_m)T/\hbar} 
\int \frac{d^f\! q'}{(2 \pi \hbar)^f} e^{i{\bf q}'\cdot{\bf p}/\hbar} 
\varphi_n({\bf q}-{\bf q}'/2) \varphi_m^*({\bf q}+{\bf q}'/2),\nonumber
\end{eqnarray}
where $W_{\varphi_n}({\bf q},{\bf p})$ is the Wigner distribution
associated to the energy eigenstate $\varphi_n({\bf q})$.
For non-linear systems, it turns out that the Wigner distribution
spreads from its initial ($T=0$) support in phase space until
it occupies most of the available phase space volume at some
preparation time $T_c$.
From time $T_c$ on, the small details of the Wigner distribution
will change with time, but in general its long scale structure
will remain as a stationary property.
To extract that characteristic long scale structure
we employ the time-averaged Wigner distribution
\begin{equation}
\overline{W_\psi({\bf q},{\bf p})}\,\equiv\,\lim_{\tau \to \infty} 
\frac{1}{\tau} \int_0^\tau dT \,W_\psi({\bf q},{\bf p},T)\,=\,
\sum_n |c_n|^2 W_{\varphi_n}({\bf q},{\bf p})\,,
\end{equation}
where we have taken
\begin{equation}
\lim_{\tau \to \infty} \frac{1}{\tau} \int_0^\tau dT \, 
\sum_{n\ne m}c_n c^*_m e^{-i (E_n-E_m)T/\hbar}
\int \frac{d^f \!q'}{(2 \pi \hbar)^f}e^{i{\bf q}'{\bf p}/\hbar} 
\varphi_n({\bf q}-{\bf q}'/2) \varphi_m^*({\bf q}+{\bf q}'/2)\,=\,0\,.
\end{equation}
Introducing $\overline{W_{\psi}}$ into Eq. (\ref{eq:overlapwig}), we obtain 
the time-averaged quantity
\begin{equation}\label{corr1}
\overline{C_{\psi}({\bf\delta q},{\bf\delta p})}\,=\,
\int d^f\! q\, d^f\! p \,\, e^{i({\bf p\cdot\delta q} + 
{\bf q\cdot\delta p}) /\hbar}\,\overline{W_\psi({\bf q},{\bf p})}\,,
\end{equation}
that describes the stationary properties of the overlap
between $|\psi_+\rangle$ and $|\psi_-\rangle$.
According to Eq. (\ref{corr1}), $\overline{C_{\psi}}$ can be identified 
as the generating function of all moments of 
$S\equiv{\bf p\cdot\delta q}+{\bf q\cdot\delta p}$
with respect to the distribution $\overline{W_{\psi}}$.  
Therefore a set of equations similar to 
Eq. (\ref{eq:overlapS}) and (\ref{eq:sqmodulusS}) 
can be obtained.
These equations imply that the initial decay of $\overline{C_\psi}$ 
is ruled by the fluctuation properties of $S$ at stationary 
conditions.
The action scale involved in the decay of $\overline{C_\psi}$
for small displacements $({\bf\delta q},{\bf\delta p})$
can be in general associated to any state
with preparation time longer than $T_c$.
This result follows from $\overline{W_{\psi}}$ describing properly
the long scale structure for $T>T_c$ and the discussion in Sec. III.
In the rest of the section we consider a family of quantum systems 
for which $\overline{C_\psi}$ can be obtained analytically.

\subsection{Systems described by the Berry-Voros conjecture}

We shall now pay special attention to (1) quantum systems with 
time-independent Hamiltonians and classical chaotic counterpart
and (2) regular quantum systems with particular random components in their 
potentials \cite{prigodin1995,srednicki1996}, for which the relevant 
quantities are obtained after averaging over the noise.
There are both, experimental and numerical evidences, that for 
these systems the so called Berry-Voros conjecture is valid, 
namely, that one can approximate the Wigner density associated 
to an energy eigenstate by a microcanonical density 
\cite{feingold1986,srednicki1994,alonso1996},
\begin{equation}
\label{bc}
W_{\varphi_n}({\bf q},{\bf p}) \rightarrow
\frac{1}{(2 \pi \hbar)^f}
\frac{\delta \big(E_n- H({\bf q},{\bf p})\big)}{\rho(E_n)},
\end{equation}
where 
$\rho(E_n)=\int\frac{d^f\! q d^f\! p}{(2\pi\hbar)^f}\delta\big(E_n-H({\bf q},
{\bf p})\big)$ 
is the local average density of states at energy $E_n$, 
and $H({\bf q},{\bf p})$ is the classical Hamiltonian 
associated to the quantum one 
\cite{berry_voros,mcdonald1988,hortikar1997,prigodin1995}. 
The Wigner distribution in the semiclassical limit fills 
the available phase space that corresponds to an energy shell 
of thickness of the order $\hbar$ and its amplitude fluctuates 
around the microcanonical density. Furthermore the density
function in Eq. (\ref{bc}) is just the leading approximation 
of a semiclassical expression for $W_{\varphi_n}({\bf q},{\bf p})$. 
The next to the leading terms depend on the periodic orbits of the 
classical system and take into account the
possible scars \cite{heller1984,bogomolny1988,agam1993}.
%
%%%%%%%%%%%%%%%%%%%%%%%%%%%%%%%%%%%%%%%%%%%%%%%%%%%%%%%%%%%%5
%

Replacing $W_{\varphi_n}({\bf q},{\bf p})$, implicit 
in Eq. (\ref{corr1}), by the expression in Eq. (\ref{bc}), 
it follows
\begin{equation}\label{micro}
\overline{C_{\psi}({\bf\delta q},{\bf\delta p})}^{BV}\,=\,
\sum_n |c_n|^2 \rho^{-1}(E_n)\, 
\Big \langle 
e^{i ({\bf p\cdot\delta q} + {\bf q\cdot\delta p})/\hbar}
\Big \rangle^{BV}_{\varphi_n}\,,
\end{equation}
where
\begin{equation}\label{eq:BVaverage}
\Big \langle 
e^{i ({\bf p\cdot\delta q} + {\bf q\cdot\delta p})/\hbar}
\Big \rangle^{BV}_{\varphi_n}\equiv
\int \frac{d^f\!q d^f\!p}{(2 \pi \hbar)^f}\, 
e^{i({\bf p\cdot \delta q} +{\bf q\cdot \delta p})/\hbar}\,
\delta\big(E_n-H({\bf q},{\bf p})\big)\,
\end{equation}
is the microcanonical average of $e^{i S/\hbar}$.
For a Hamiltonian of the form 
$H({\bf q},{\bf p})={\bf p}^2/2M + V({\bf q})$,
and after integrating over the momentum variables, 
one obtains
\begin{eqnarray}
\label{Wf}
\Big\langle 
e^{i ({\bf p\cdot\delta q} + {\bf q\cdot\delta p})/\hbar} 
\Big\rangle^{BV}_{\varphi_n}
&=& (2 \pi)^{f/2} M \int   \frac{d^f\!q}{(2 \pi \hbar)^f} 
e^{i {\bf q\cdot\delta p}/\hbar}
\Big( \frac{\hbar}{|{\bf\delta q}|} \sqrt{2M(E_n-V({\bf q}))}\Big)^{\frac{f}{2}-1} 
\cr \nonumber\\
&&\times 
J_{\frac{f}{2} -1}\Big(\frac{|{\bf\delta q}|}{\hbar} \sqrt{2M(E_n-V({\bf q}))}\Big),
\end{eqnarray}
where $J_{\frac{f}{2} -1}(z)$ is the Bessel function 
of order $f/2-1$. Eqs. (\ref{micro}) and (\ref{Wf}) lead 
to a formal expression of the time-averaged two-point correlation
function $\overline{C_{\psi}({\bf\delta q},{\bf\delta p})}^{BV}$ in terms 
of the potential $V({\bf q})$.
These equations constitute the main result of this section and
are the starting point for the analysis of particular examples.
In the following we shall particularise Eq. (\ref{micro}) for systems with
a random component in the potential such that the average over the noise is
the $f$-dimensional harmonic potential.

\subsubsection{The $f$-dimensional harmonic oscillator}

The classical Hamiltonian for a generic $f$-dimensional harmonic oscillator,
\begin{equation}
H(\tilde{\bf q},\tilde{\bf p})\,=\,
\sum_{i=1}^{f}\,\frac{\tilde{p}_i^2}{2m_i}
\,+\,\frac{1}{2}m_i \omega_i^2 \tilde{q}_i^2\,,
\end{equation}
can be rewritten, in terms of the rescaled coordinates and momenta
\begin{eqnarray}
p_i&\equiv&\sqrt{\frac{M}{m_i}}\,\tilde{p}_i\nonumber\\
q_i&\equiv&\sqrt{\frac{m_i\omega_i^2}{M\omega^2}}\,\tilde{q}_i\,,
\end{eqnarray}
as the spherical harmonic oscillator
\begin{equation}
H({\bf q},{\bf p})\,=\,\frac{1}{2M} (p_1^2+p_2^2+\cdots+p_f^2)+
\frac{1}{2}M \omega^2 (q_1^2+q_2^2+\cdots+q_f^2)\,.
\end{equation}
The integral in Eq. (\ref{eq:BVaverage}) reads
\begin{equation}
\left(\prod_{i=1}^{f}\frac{\omega}{\omega_i}\right)
\int \frac{d^f\!q\, d^f\!p}{(2 \pi \hbar)^f}\, 
e^{i({\bf p\cdot \delta q} +{\bf q\cdot \delta p})/\hbar}\,
\delta\big(E_n-H({\bf q},{\bf p})\big)\,,
\end{equation}
with
\begin{eqnarray}
\delta {\bf q}&\equiv&
\left(\sqrt{\frac{m_1}{M}}\,\delta \tilde{q}_1,\dots,
\sqrt{\frac{m_f}{M}}\,\delta \tilde{q}_f\right)\nonumber\\
\delta {\bf p}&\equiv&
\left(\sqrt{\frac{M\omega^2}{m_1\omega_1^2}}\,\delta \tilde{p}_1,\dots,
\sqrt{\frac{M\omega^2}{m_f\omega_f^2}}\,\delta \tilde{p}_f\right)\,.
\end{eqnarray}
After some manipulations, it follows that
\begin{eqnarray}
&&\Big \langle 
e^{i ({\bf p\cdot \delta q} + {\bf q\cdot \delta p})/\hbar} 
\Big \rangle^{BV}_{\varphi_n}\,=\,
2^{f-1}\,\left(\frac{\omega^f}{\prod_{i=1}^f\,\omega_i}\right) 
\frac{E_n^{f-1}}{(\hbar \omega)^f}
\Big( \frac{\hbar}{|{\bf \delta q} |} 
\sqrt{\frac{1}{2 M E_n}}\Big)^{\frac{f}{2}-1}
\Big( \frac{\hbar}{|{\bf \delta p} |} 
\sqrt{\frac{M \omega^2}{2 E_n}}\Big)^{\frac{f}{2}-1}\nonumber\\ 
&&\qquad \times \int_0^1 d \,\xi \,
\xi^{\frac{f}{2}} \Big(\sqrt{1-\xi^2}\Big)^{\frac{f}{2}-1}
J_{\frac{f}{2}-1}\Big( \frac{{|\bf\delta q}|}{\hbar} \sqrt{2ME_n} 
\sqrt{1-\xi^2}\Big)J_{\frac{f}{2}-1}\Big(\frac{|{\bf\delta p}|}{\hbar}
\xi\sqrt{\frac{2E_n}{M \omega^2}} \Big)\,,
\end{eqnarray}
and integrating over variable $\xi$,
\begin{equation}
\Big \langle 
e^{i ({\bf p\cdot \delta q} + {\bf q\cdot \delta p})/\hbar} 
\Big \rangle^{BV}_{\varphi_n}\,=\,
2^{f-1} \,\left(\frac{\omega^f}{\prod_{i=1}^f\,\omega_i}\right)
\frac{E_n^{f-1}}{(\hbar \omega)^f}
\frac{J_{f-1} \Big( 
\sqrt{(\frac{|{\bf\delta p}|}{\hbar} \sqrt{\frac{2E}{M \omega^2}})^2+
(\frac{|{\bf\delta q}|}{\hbar} \sqrt{2M E})^2 }   \Big)}
{ \sqrt{\Big((\frac{|{\bf\delta p}|}{\hbar} \sqrt{\frac{2E}{M \omega^2}})^2+
(\frac{|{\bf\delta q}|}{\hbar} \sqrt{2M E})^2 \Big)^{f-1}} }\,.
\end{equation}
For an eigenstate of energy $E_n$,
$\rho(E_n)= E_n^{f-1}/(\Gamma(f)(\hbar \omega)^f)$, 
where $\Gamma(f)$ denotes the Gamma function of argument $f$.
Besides, $\sigma_{p,n}^2=\la{\bf p}^2\ra^{BV}_{\varphi_n}= ME_n$,
$\sigma_{q,n}^2=\la{\bf q}^2\ra^{BV}_{\varphi_n}= E_n/M\omega^2$ 
(for this case the mean values of position and of momentum vanish), 
giving
\begin{eqnarray}\label{eq:bessel}
\Big \langle 
e^{i ({\bf p\cdot \delta q} +{\bf q\cdot \delta p})/\hbar} 
\Big \rangle^{BV}_{\varphi_n}
&=&
{2}^{f-1} \,\left(\frac{\omega^f}{\prod_{i=1}^f\,\omega_i}\right) 
\Gamma (f)\rho(E_n) \frac{J_{f-1} \Big(\sqrt{2}
\sqrt{(\frac{|{\bf\delta p}| \sigma_{q,n}}{\hbar} )^2+
      (\frac{|{\bf\delta q}| \sigma_{p,n}}{\hbar} )^2 }   \Big)}
{\sqrt{((\frac{|{\bf\delta p} |\sigma_{q,n}}{\hbar})^2+
                (\frac{|{\bf\delta q}|\sigma_{p,n}}{\hbar})^2 )^{f-1}}}\nonumber\\
&=&{2}^{f-1} \,\left(\frac{\omega^f}{\prod_{i=1}^f\,\omega_i}\right)
\Gamma (f) \rho(E_n)
\frac{J_{f-1}(\sqrt{2} {\Delta {S}^{BV}_n}/\hbar)}
{( \Delta {S}^{BV}_n/\hbar)^{f-1}}\,,
\end{eqnarray}
where the characteristic action
\begin{equation}\label{eq:actionsn}
\Delta {S}^{BV}_n=
\sqrt{(|{\bf\delta p}|\sigma_{q,n})^2+(|{\bf\delta q}|\sigma_{p,n})^2}
\end{equation}
has been introduced. 
$\Delta S^{BV}_n$ is nothing but
the action $\Delta S$ introduced in Eq. (\ref{eq:completa}) 
calculated for the $n$th eigenstate using the Berry-Voros 
conjecture.
For the superposition state (\ref{psi}), one obtains
\begin{equation}\label{eq:meanC}
\overline{ C_{\psi}({\bf\delta q},{\bf\delta p})}^{BV} = 
\sum_n {2}^{(f-1)/2} \,
|c_n|^2\, \Gamma (f) 
\frac{J_{f-1}(\sqrt{2} \Delta {S}^{BV}_n/\hbar)}
{(\Delta {S}^{BV}_n/\hbar)^{f-1}}\,,
\end{equation}
so that the typical action that controls the decay of the overlap 
is the one related with the coefficients $c_n$ that contribute 
more to the initial state. The action $\overline{\Delta S}^{BV}$,
evaluated for the average distribution $\overline{W_{\psi}}$ under
the Berry-Voros conjecture, is related with $(\Delta S^{BV}_n)$ by
\begin{equation}
(\overline{\Delta S}^{BV})^2\,=\,\sum_n |c_n|^2 (\Delta S^{BV}_n)^2\,.
\end{equation}
For any displacement $({\bf\delta q},{\bf\delta p})$ the 
previous relation can be inverted and used to write 
$\overline{ C_{\psi}}^{BV}$ in terms of $\overline{\Delta S}^{BV}$.

\begin{figure}
\includegraphics[width=10cm]{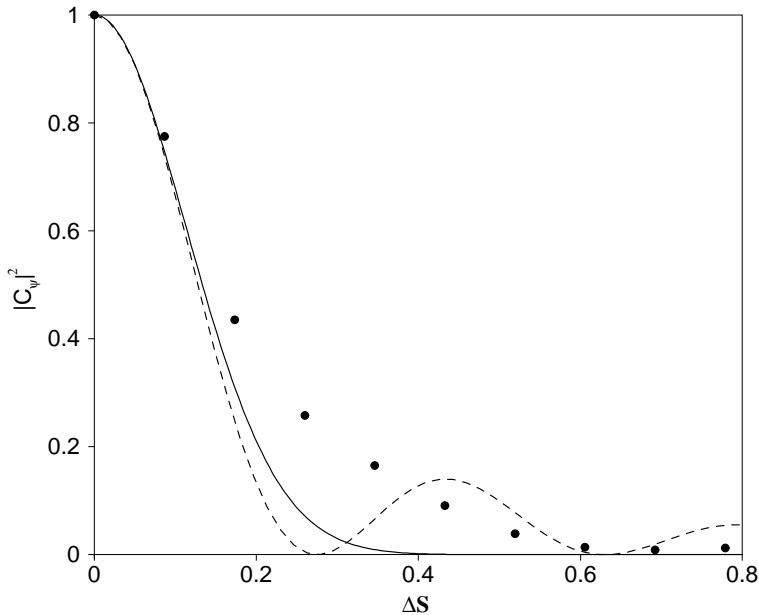}
\vskip0.5cm
\caption[]{$|\overline{C_{\psi}}^{BV}|^2$ versus $\overline{\Delta S}^{BV}$ 
(dashed line) for the direction $\delta q\simeq 6.8 \delta p$ in phase
space and the same initial state as in Fig. \ref{fig:fig1}. 
For comparison, $|{C}_{\psi}|^2$ is shown for the initial (solid line)
and $T=500$ (dots) states of Fig. \ref{fig:fig2}. Arbitrary units are
used.}
\label{fig:fig3}
\end{figure}

To illustrate this result, in Fig. \ref{fig:fig3} we plot 
$\overline{C_{\psi}}^{BV}$ versus $\overline{\Delta S}^{BV}$ for the 
one-dimensional case $f=1$. 
%
%%%%%%%%%%%%%%%%%%%%%%%%%%%%%%
%%%%%%%%%%%%%%%%%%%%%%%%%%%%%%%
%
The $\Delta S$-action scale for the decay of the overlap is dictated by 
the value $\hbar=0.16$, as in the case described in the previous 
section.
Note that this result is expected as a power expansion of the 
Bessel function $J_0$ in Eq. (\ref{eq:meanC}) to the second order 
in $\overline{\Delta S}^{BV}$ consistently recovers the result in 
Eq. (\ref{eq:sqmodulusS}).

%%%%%%%%%%%%%%%%%%%%%%%%%%%%%%%%%%%%%%%%%%%%%%%%%%%%%%%%%%%%%%%%%%%%%%%%%%

\section{Discussion}

The results of previous sections show that the relevant
$\Delta S$-action scale to the decay
of the overlap $|C_\psi|^2$ for small displacements 
$(\delta {\bf q},\delta {\bf p})$ is given by $\hbar$.
The one-dimensional Gaussian state is a
special example for which the dependence of $|C_\psi|^2$ 
on $\Delta S$ is given explicitly by Eq. (\ref{eq:cforgaussian}),
and its monotonic exponential decay is independent of 
particular details of the state, 
as for instance the widths in position and momentum. 
(On the contrary, the decay of the overlap with the 
displacement will depend on $\sigma_q$ and $\sigma_p$ 
through Eq. (\ref{eq:dsforgaussian})). 
In Figures \ref{fig:fig2} and \ref{fig:fig3}
the exponential dependence associated to 
an initial (Gaussian) coherent state 
is compared to the one corresponding to 
states at different preparation times.
Although all the curves shows a similar initial decay,
(dictated by $\Delta S\approx\hbar$),
the ulterior behaviour can have qualitatively different features,
the presence of oscillations in the overlap for intermediate 
values of $\Delta S$ being the most relevant one.
It is worth noting that these oscillations can never be regarded
as true revivals. $|C_\psi|^2$ can be interpreted as 
the overlap between the states  $|\psi\rangle$ and 
$\hat{D}(\delta{\bf q},\delta{\bf p})|\psi\rangle$, 
the second one being obtained by a rigid displacement of 
$|\psi\rangle$. This implies that $|C_\psi|^2$ can not be
equal to one for non-zero displacements since the support
of the state in phase-space in finite.
However, large amplitude oscillations are possible
as shown in Fig. \ref{fig:fig2} for $T=10$.

The pattern of oscillations will change in general
with the preparation time.
In the system described in Sec. \ref{sec:1dtd}, 
no oscillations are present for the initial state.
For small preparation times some oscillations appear
(see Fig. \ref{fig:fig2} for $T=10$)
but their amplitude decrease when the preparation time increases.
For larger preparation times only oscillations with small amplitude 
are found.
This behaviour can be interpreted by using Eq. 
(\ref{eq:scalar_product_k1}) with
the Wigner function, as proposed in Ref. \cite{zurek2001}. 
For $T=0$, the Wigner distribution
associated to the coherent initial state is a Gaussian in phase space,
and the monotonic decrease of $|C_ \psi|^2$ with 
$\Delta S$ reflects the decrease of the overlapping regions
between the states $|\psi_+\rangle$ and $|\psi_-\rangle$
(or, equivalently, between $|\psi\rangle$ and $\hat{D}|\psi\rangle$).
For small preparation times, the isolated evolution of the
environmental system prior to the coupling generates a regular 
large scale structure in the distribution (characterised by large values 
of $\Delta Z_0$ in Fig \ref{fig:fig2}). 
For this case
the coincidence between maxima and minima of that 
large scale structure in $|\psi_+\rangle$ and $|\psi_-\rangle$
is responsible for the oscillations in the overlap.
For longer preparation times, smaller scale structures
appear in the distribution (corresponding to smaller values 
of $\Delta Z_0$), and more importantly, the randomness of 
the distribution of the patches in the structure increases 
(reflected in the similarity of the actions $\Delta Z_0$ and $a$).
Then, as $T$ increases the amplitude of the oscillations 
becomes smaller until they are eventually negligible.
This behaviour is expected in general for 
any non-linear system, with the only difference in 
the preparation time $T$ needed to develop the small scale
structure.

In the light of this discussion, 
special care must be taken in the interpretation
of the results of Sec. \ref{sec:nonlinear}, where
broad oscillations in the time averaged overlap 
$|\overline{C_\psi}^{BV}|^2$ could appear for large
$\overline{\Delta S}^{BV}$ (see Fig. \ref{fig:fig3}).
As the systems considered are non-linear, the states will develop
in general a complex small random structure for long enough 
preparation times, and only negligible oscillations will 
be present in the overlap.
The broad oscillations in $|\overline{C_\psi}^{BV}|^2$ 
are the result of the use of the Berry-Voros conjecture, 
that describes correctly the large scale structure
but fails in describing the small scale correlations.
Therefore, following the discussion in Sec. \ref{sec:structures} 
related to Eq. (\ref{Uni}), only the initial decay (corresponding 
to small displacements) for each particular sufficiently long
preparation time is well described by $|\overline{C_\psi}^{BV}|^2$.
In the approach used in this work, the effect of the coupled 
evolution in the environmental system is equivalent to rigid 
displacements in phase-space of the state $|\psi(T)\rangle$
to give $|\psi_+(T;\delta t)\rangle$ and $|\psi_-(T;\delta t)\rangle$. 
(The dependence of $|\psi_\pm\rangle$ with the interaction time
$\delta t$ is made explicit.)
No additional structure in phase-space in the states 
$|\psi_\pm\rangle$ is generated during the coupling, as 
the contribution of $\hat{H}_{\cal E}$ is neglected.
The interaction time $\delta t_0$ required to
obtain a value $|C_0|^2$ of the overlap is given
by the condition 
$|C_\psi(\delta{\bf q}_0,\delta{\bf p}_0)|^2=|C_0|^2$,
where the magnitude of the displacements are 
${\bf \delta q}_0=-2{\bf c_p}\delta t_0$ and
${\bf \delta p}_0=-2{\bf c_q}\delta t_0$.
Therefore the larger the coupling constants, the smaller 
the interaction time $\delta t_0$.
The condition $\Delta S\approx \hbar$ establishes a lower bound
for the value of the displacements and consequently for the 
interaction time needed to attain effective decoherence.
An alternative derivation of the lower bound is pointed
out in Ref. \cite{KJZ02}.
A different aspect is the dependence of this $\delta t_0$
with the environmental state prior to the interaction.
The size of the displacements $(\delta{\bf q}_0,\delta{\bf p}_0)$
can be described by the action $\Delta Z_0=\delta{\bf q}_0\delta{\bf p}_0$ 
for each particular state. 
As discussed in Sec. \ref{sec:structures}, $\Delta Z_0$ is
of the order of the action that sets the scale of the structures
in the distribution for some phase space 
representations fulfilling Eq. (\ref{eq:scalar_product_k1}).
As a result, $\delta t_0$ decreases as the structure in
the distribution associated to the state becomes smaller.
For example, in the system analysed in Fig \ref{fig:fig1},
the interaction time $\delta t_0$ is proportional 
to $\sqrt{a}$ \cite{KJZ02},
provided the preparation time is long enough for $a$ 
to describe properly the small scale structure.
A more complex situation appears when the evolution induced
by $\hat{H}_{\cal E}$ is not neglected \cite{KJZ02,commentprl}.
In that case, besides the displacement, the distribution of the
structure in phase space of the states $|\psi_+\rangle$ and 
$|\psi_-\rangle$ will change during the interaction. 
For the system studied in Fig \ref{fig:fig1}
two different regimes can be distinguished. 
For $T\lesssim 20$, a rapid variation of the sizes of the structure
with time is found and both mechanisms, the displacement and 
the development of structure, will determine the interaction
time $\delta t_0$.
However, for $T\gtrsim 20$, the variation of 
the sizes of the structure is much slower and $\delta t_0$ is 
determined by the time required to produce the displacement 
in phase space. As the displacement is approximately 
independent of the details of the state, $\delta t_0$ will be 
weakly dependent on the preparation time for $T\gtrsim 20$ 
\cite{KJZ02}.
Another important point to discuss is the dependence 
of the decoherence process with the number of degrees 
of freedom of the environmental system.
As the number of degrees of freedom increases,
smaller displacements in each variable are needed to 
obtain $\Delta S\approx\hbar$, that sets the action scale 
for the initial decay of the overlap in all cases, and
the corresponding interaction time will be smaller too.
This is compatible with the observation that the larger 
the environment the more effective the decoherence process.

Experimental tests of the decoherence process in the context
discussed in this work can be in principle realized in 
the systems described in Refs. \cite{exper}. 
The interaction between two oscillators is mediated by a term 
of the form 
%$\hbar G a_^\dagger a(b+b^\dagger)$, 
$\hbar\,G\,a_{\cal S}^\dagger\,a_{\cal S}\,(a_{\cal E}+a_{\cal E}^\dagger)$,
corresponding to
a scattering process in which a quantum of energy of 
the environmental system $\cal E$ can be absorbed ($a_{\cal E}$) 
or emitted ($a_{\cal E}^\dagger$) whereas the
number of quanta of the pointer system $\cal S$ 
remains the same.
For these cases, the coherences of the reduced density operator
of the pointer system in the basis given by the Fock states are
proportional to the overlap between the states
$\hat D(\alpha=iGn\delta t)|\psi(T)\rangle$ and
$\hat D(\alpha'=iGn'\delta t)|\psi(T)\rangle$.
The operator
$\hat D(\alpha=iGn\delta t)\equiv
\exp\{\alpha a_{\cal E}^\dagger-\alpha^*a_{\cal E}\}$
produces a displacement in phase space that depends
linearly on the interaction time $\delta t$, the coupling constant 
$G$, and the index $n$ of one of the Fock states of $\cal S$ 
involved in the coherence under consideration.

In summary, the role of $\hbar$ as a boundary between different 
decoherence regimes has been clarified in the context of 
a characteristic action $\Delta S$, which depends on the 
quantum state of the environmental system. 
We related the action $\Delta S$ with the complementary
quantity $\Delta Z$, and described their connection with
the pattern of structures developed in phase space.

%%%%%%%%%%%%%%%%%%%%%%%%%%%%%%%%%%%%%%%%%%%%%%%%%%%%%%%%%%%%%%%%%%%%%%%%%%
\acknowledgments{
We thank W. Zurek and the referee for useful comments. 
This work was supported by the 
``Ministerio de Ciencia y Tecnolog\'\i a'' under FEDER BFM2001-3349, by
``Consejer\'\i a de Educaci\'on Cultura y Deportes (Gobierno de Canarias)''
under Contract No. PI2002-009, and by CERION II (Canadian European Research
Initiative on Nanostructures).}

%%%%%%%%%%%%%%%%%%%%%%%%%%%%%%%%%%%%%%%%%%%%%%%%%%%%%%%%%%%%%%


\begin{thebibliography}{8.}
\bibitem{commentprl} A different approach is used in 
Ref. \cite{KJZ02}, where the evolution under Hamiltonian 
(\ref{eq:HamZurek}) is not neglected during the interaction 
process, and the coupling term considered results in a shift
of the position of the minimum of the confining harmonic potential.
\end{thebibliography}
\end{document}